\documentclass[aps, prd, floatfix, nofootinbib, showpacs, twocolumn, superscriptaddress]{revtex4-1}

\usepackage{amsmath,amsfonts,amssymb,bm}
\usepackage{graphicx}
\usepackage{color}
\usepackage{subfigure}
\usepackage{multirow}
\usepackage{textcomp}
\usepackage{slashed}

\usepackage[mathscr,scaled=1.15]{urwchancal}
\DeclareFontFamily{OT1}{pzc}{}
\DeclareFontShape{OT1}{pzc}{m}{it}%
{<-> s * [1.15] pzcmi7t}{}
\DeclareMathAlphabet{\mathpzc}{OT1}{pzc}{m}{it}

\definecolor{purple}{rgb}{0.5,0,0.5}
\definecolor{blue}{rgb}{0.0,0,0.9}
\definecolor{prdblue}{rgb}{0.133,0.118,0.498}
\usepackage[colorlinks=true, pdfstartview=FitV, linkcolor=prdblue, citecolor= prdblue, urlcolor=prdblue]{hyperref}

\begin{document}

\title{$\,$\\[-7ex]\hspace*{\fill}{\normalsize{\sf\emph{Preprint no}. NJU-INP 018/20}}\\[1ex]
Reflections upon the Emergence of Hadronic Mass}

\author{Craig D.~Roberts}
\email[]{cdroberts@nju.edu.cn}
\affiliation{School of Physics, Nanjing University, Nanjing, Jiangsu 210093, China}
\affiliation{Institute for Nonperturbative Physics, Nanjing University, Nanjing, Jiangsu 210093, China}

\author{Sebastian M.~Schmidt}
\email[]{s.schmidt@hzdr.de}
\affiliation{RWTH Aachen University, III. Physikalisches Institut B, Aachen D-52074, Germany}
\affiliation{Helmholtz-Zentrum Dresden-Rossendorf, Dresden D-01314, Germany}

\date{25 May 2020}

\begin{abstract}
\hspace*{-\parindent}\mbox{\sf Abstract}.
With discovery of the Higgs boson, science has located the source for $\lesssim 2$\% of the mass of visible matter.  The focus of attention can now shift to the search for the origin of the remaining $\gtrsim 98$\%.  The instruments at work here must be capable of simultaneously generating the 1\,GeV mass-scale associated with the nucleon and ensuring that this mass-scale is completely hidden in the chiral-limit pion.  This hunt for an understanding of the emergence of hadronic mass (EHM) has actually been underway for many years.  What is changing are the impacts of QCD-related theory, through the elucidation of clear signals for EHM in hadron observables, and the ability of modern and planned experimental facilities to access these observables.  These developments are exemplified in a discussion of the evolving understanding of pion and kaon parton distributions.
\end{abstract}

\maketitle

\section{Introduction}
A key question posed to modern science centres on the character of mass and its consequences in the Standard Model, especially as it emerges from the strong interaction sector; namely, quantum chromodynamics (QCD).  When considering the source of mass, the first thing that comes to many minds is explicit mass generation via the Higgs-mechanism; especially because the Higgs boson was discovered relatively recently \cite{Aad:2012tfa, Chatrchyan:2012xdj}, with its importance acknowledged by the subsequent Nobel Prize awarded to Englert and Higgs \cite{Englert:2014zpa, Higgs:2014aqa}: ``\emph{for the theoretical discovery of a mechanism that contributes to our understanding of the origin of mass of subatomic particles} \ldots''.

\begin{figure}[b]
\centerline{%
\includegraphics[clip, width=0.475\textwidth]{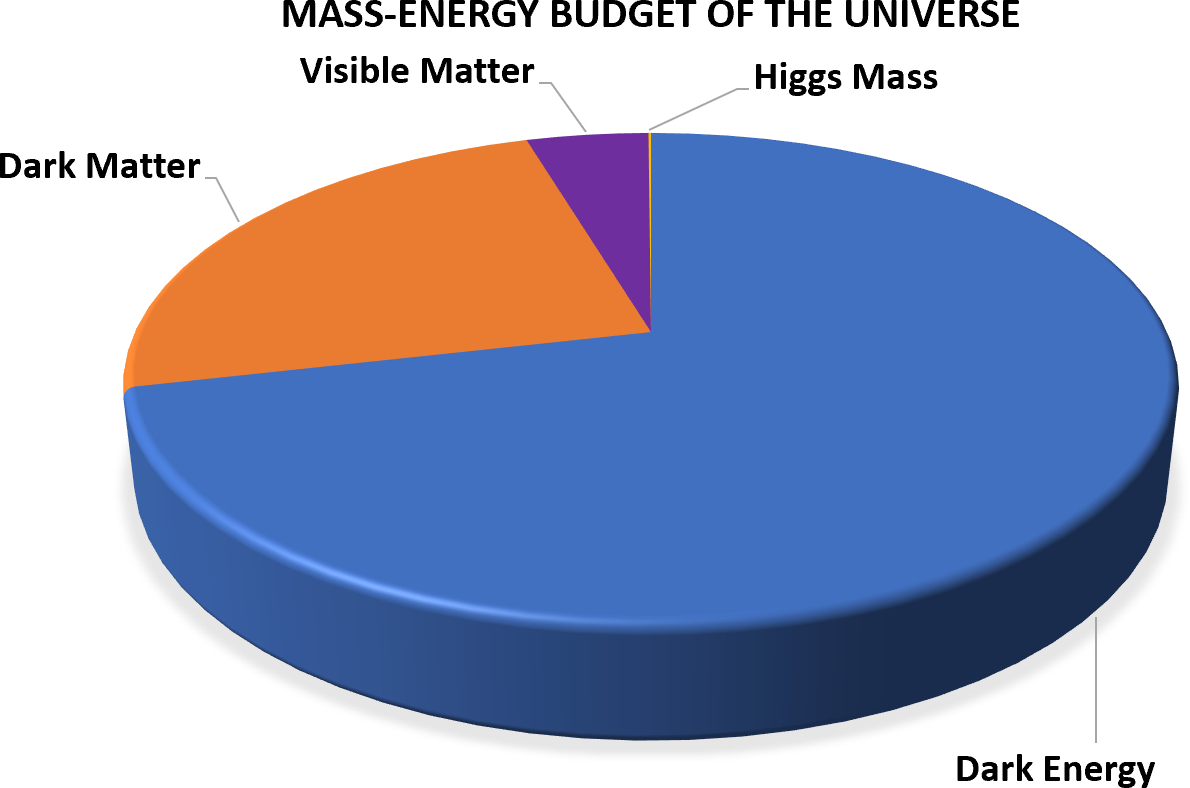}}
\caption{\label{FigMassEnergy}
The Wilkinson Microwave Anisotropy Probe (WMAP) \cite{limon/etal:2003} determined that the universe is flat, from which it follows that the mean energy density in the Universe is equal to the critical density, \emph{viz}.\ only 5.9 protons/m$^3$.  Of this total density \cite{Seife2038}: dark energy is 71\%; dark matter is another 24\%; and the remaining 5\% is visible material.  Of this 5\%, less than 0.1\% is tied simply to the Higgs boson, as indicated by the yellow sliver at the top of the disk.
}
\end{figure}

Discovery of the Higgs was a watershed.  However, it should be placed in context.  Therefore, consider the mass-energy budget of the Universe, illustrated in Fig.\,\ref{FigMassEnergy}: dark energy constitutes 71\%; dark matter is another 24\%; and the remaining 5\% is visible material.  Little is known about the first two: science can currently say almost nothing about 95\% of the mass-energy in the Universe.  The explanation lies outside the Standard Model.  On the other hand, the remaining 5\% has forever been the source of everything tangible.  Yet, amongst this 5\%, less-than 0.1\% is tied directly to the Higgs boson; hence, even concerning visible material, too much remains unknown.

More than 98\% of visible mass is contained within nuclei.  In first approximation, their atomic weights are simply the sum of the masses of all the neutrons and protons (nucleons) they contain.  Each nucleon has a mass $m_N \sim 1\,$GeV, \emph{i.e}.\ approximately 2000-times the electron mass.  The Higgs boson produces the latter, but what produces the masses of the neutron and proton?  This is the question posed above, which is pivotal to the development of modern physics: how can science explain the emergence of hadronic mass (EHM)?

Within the Standard Model, strong interactions are described by QCD.  In this theory, nucleons and all similar objects (hadrons), are composites, built from quarks and/or antiquarks (matter/antimatter fields), held together by forces produced by the exchange of gluons (gauge fields).  These forces are unlike any previously encountered.  Extraordinarily, \emph{e.g}.\ they become very weak when two quarks are brought close together within a nucleon \cite{Politzer:2005kc, Wilczek:2005az, Gross:2005kv}.  However, all attempts to remove a single quark from within a nucleon and isolate it in a detector have failed.  Seemingly, then, the forces become enormously strong as the separation between quarks is increased \cite{Wilson:1974sk}.

Modern science is thus encumbered with the fundamental problem of gluon and quark confinement; and confinement is crucial because it ensures absolute stability of the proton.  In the absence of confinement, protons in isolation could decay; the hydrogen atom would be unstable; nucleosynthesis would be a chance event, having no lasting consequences; and without nuclei, there would be no stars and no living Universe.  Without confinement, our Universe cannot exist.

As the 21st Century began, the Clay Mathematics Institute established seven Millennium Prize Problems \cite{millennium:2006}.  Each represents one of the toughest challenges in mathematics.  The set contains the problem of confinement; and presenting a sound solution will win its discoverer \$1,000,000.  Even with such motivation, today, almost fifty years after the discovery of quarks \cite{Taylor:1991ew, Kendall:1991np, Friedman:1991nq}, no rigorous solution has been found.  Confinement and EHM are inextricably linked. Consequently, as science plans for the next thirty years, solving the problem of EHM has become a \emph{grand challenge}.

\section{QCD is Full of Surprises}
\begin{figure}[t]
\vspace*{1ex}

\centerline{%
\includegraphics[clip, width=0.475\textwidth]{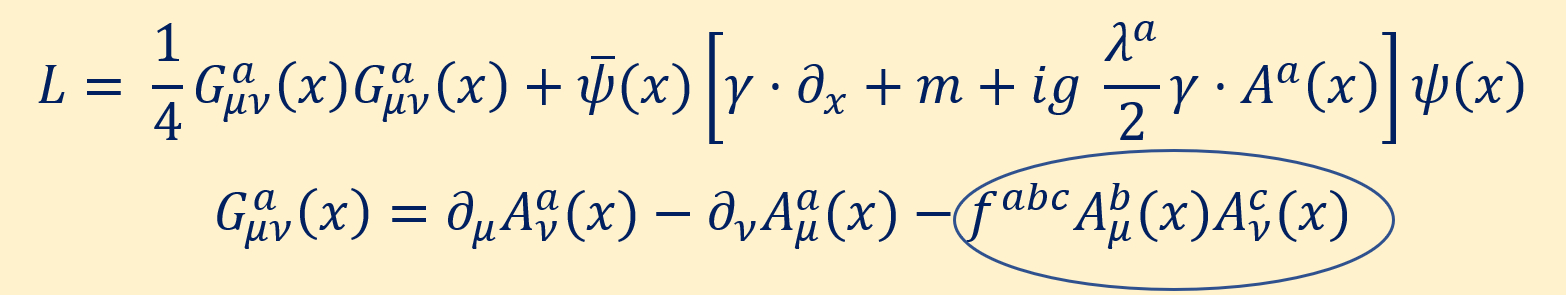}}
\caption{\label{FigQCD}
QCD's Lagrangian density can be expressed in merely two lines: $\psi$ -- quark field; $\bar\psi$ -- antiquark; and $A_\mu^a$, $a=1,\ldots,8$ -- gluon field.  The matrices $\{\tfrac{1}{2} \lambda^a\}$ are the generators of the SU$(3)$ (color/chromo) gauge-group in the fundamental representation; and $m$ is the current-quark mass generated by the Higgs boson.
}
\end{figure}

Regarding Fig.\,\ref{FigQCD}, in appearance, QCD is simple.  It is also unique.  QCD is a fundamental theory with the capacity to sustain massless elementary degrees-of-freedom, \emph{viz}.\ gluons and quarks; yet gluons and quarks are predicted to acquire mass dynamically \cite{Lane:1974he, Politzer:1976tv, Pagels:1978ba, Cornwall:1981zr, Higashijima:1983gx, Munczek:1983dx, Fomin:1984tv, Cahill:1985mh, Aguilar:2015bud}, and nucleons and almost all other hadrons likewise, so that the only (nearly-) massless systems in QCD are its composite (pseudo-) Nambu-Goldstone (NG) bosons \cite{Nambu:1960tm, Goldstone:1961eq, Horn:2016rip}, \emph{e.g}.\ pions and kaons.

The existence of pions and kaons and their ``unnaturally'' small masses are empirical facts \cite{Tanabashi:2018oca}.  Following their discovery \cite{Lattes:1947mw, Rochester:1947mi}, identification of $\pi$- and $K$-mesons as the NG modes that emerge with dynamical chiral symmetry breaking (DCSB -- explained below) in the Standard Model was a gradual process \cite{Marciano:1977su}.  Elucidation of the diverse array of local and global consequences of this association is still underway.  Aspects of the ongoing effort are sketched herein, with a focus on parton distribution functions (PDFs), the study of which has a long history.  Introduced fifty years ago \cite{Bjorken:1969ja}, PDFs are probability densities that catalogue the sharing of momentum between all participants in a fully relativistic quantum field theory bound state.  Notably, today, sound QCD-connected predictions are at last becoming available.


An understanding of QCD's NG modes is basic to any solution of the Standard Model; yet, they are peculiar.  For example, pions and kaons are responsible for binding systems as diverse as atomic nuclei and neutron stars, but the energy associated with the gluons and quarks within these NG modes is not readily apparent.  This contrasts starkly with all other ``everyday'' hadronic bound states, \emph{viz}.\  systems constituted from \mbox{$u$-,} $d$-, and/or $s$-quarks, which possess nuclear-size masses far in excess of anything that can directly be tied to the Higgs boson.

In trying to match QCD with Nature, one confronts the many complexities of strong, nonlinear dynamics in relativistic quantum field theory, \emph{e.g}.\ the loss of particle number conservation, the frame and scale dependence of the explanations and interpretations of observable processes, and the evolving character of the relevant degrees-of-freedom. Electroweak theory and phenomena are essentially perturbative; hence, possess little of this complexity. Science has never before encountered an interaction such as that at work in QCD. Understanding this interaction, explaining everything of which it is capable, can potentially change the way we look at the Universe.

In comparison with quantum electrodynamics (QED), the sole, essential difference is the circled term in Fig.\,\ref{FigQCD}, describing gluon self-interactions.  If QCD is correct, a conjecture supported by its ability to describe a wide variety of high-energy phenomena \cite{Tanabashi:2018oca}, then this term must hold the answers to an enormous number of Nature's basic questions, \emph{e.g}.: what is the origin of visible mass and how is it distributed within atomic nuclei; and what carries the proton's spin and how can the same degrees-of-freedom combine to ensure the pion is spinless?  Nowhere in the visible Universe are there more basic expressions of \emph{emergence}.

Treated as a classical theory, chromodynamics is a non-Abelian local gauge field theory.  Formulated in four spacetime dimensions, such theories do not possess any mass-scale in the absence of Lagrangian masses for the quarks. There is no dynamics in a scale-invariant theory, only kinematics. Bound states are therefore impossible and, accordingly, our Universe cannot exist \cite{Roberts:2016vyn}.  A spontaneous breaking of symmetry, as realized via the Higgs mechanism, does not solve this problem: the masses of the neutron and proton, the kernels of all visible matter, are roughly 100-times larger than the Higgs-generated current-masses of the light $u$- and $d$-quarks, the valence constituents of nucleons.

On the flip side, the real world's composite NG bosons are (nearly) massless.  Hence, in these systems, the strong interaction's $m_N := m_{\rm proton} \approx 1\,$GeV mass-scale is effectively hidden. In fact, there is a particular circumstance in which the pseudoscalar mesons $\pi$, $K$, $\eta$ are exactly massless, \emph{i.e}.\ the chiral limit, when the Higgs-generated masses in Fig.\,\ref{FigQCD} are omitted.  In this case, perturbative QCD predicts that strong interactions cannot distinguish between quarks with negative or positive helicity. Such a chiral symmetry would have numerous corollaries, \emph{e.g}.\ existence of a scalar meson degenerate with the pion.  However, no state of this type is observed.  In fact, none of the consequences of this chiral symmetry are found in Nature.  Instead, the symmetry is broken by interactions.  DCSB is the agent behind both the massless quarks in QCD's Lagrangian acquiring a large effective mass \cite{Bhagwat:2003vw, Bowman:2005vx, Bhagwat:2006tu} and the interaction energy between those quarks cancelling the sum of their masses exactly so that the composite pion is massless in the chiral limit \cite{Maris:1997hd, Qin:2014vya, Binosi:2016rxz}.

Reinstating the Higgs mechanism, then, as illustrated elsewhere (\emph{e.g}.\ Refs.\,\cite{Bender:1996bb, Maris:1997tm, Eichmann:2009qa}), DCSB is responsible for, \emph{inter alia}: the physical size of the pion mass ($m_\pi \approx 0.15 \,m_N$); the large mass-splitting between the pion and its valence-quark spin-flip partner, the $\rho$-meson ($m_\rho > 5 \,m_\pi$); and the neutron and proton possessing masses $m_N \approx 1\,$GeV. Interesting things happen to the kaon, too. Like a pion, but with one of the light quarks replaced by a $s$-quark, the kaon comes to possess a mass $m_K \approx 0.5\,$GeV. Here a competition is taking place, between dynamical and Higgs-driven mass generation.

These phenomena and features, their origins and corollaries, entail that the question of how did the Universe evolve is inseparable from the questions of what is the source of the $m_N \approx 1\,$GeV mass-scale that characterizes atomic nuclei; why does $m_N$ have the observed value; and, enigmatically, why does the dynamical generation of $m_N$ have seemingly no effect on QCD's composite NG bosons, \emph{i.e}.\ whence the near-absence of the pion mass?

\section{QCD's Running Coupling}
When considering confinement, the definition is a core issue.  Ask a practitioner and one will receive an answer; yet the perspectives of any two people are often distinct, \emph{e.g}.\ Refs.\,\cite{Wilson:1974sk, Gribov:1998kb, Cornwall:1981zr}.  The proof of one expression of confinement will be contained within a demonstration that quantum SU$_c(3)$ gauge field theory is mathematically well-defined, \emph{i.e}.\ a solution to the ``Millennium Problem'' \cite{millennium:2006}.  However, that may be of limited value because Nature has provided light-quark degrees-of-freedom, which seemingly play a crucial r\^ole in the empirical realisation of confinement, perhaps because they enable screening of colour charge at low coupling strengths \cite{Gribov:1998kb}.

QCD's running coupling is basic to most attempts to define and understand confinement because, almost immediately following the demonstration of asymptotic freedom \cite{Politzer:2005kc, Wilczek:2005az, Gross:2005kv}, the associated appearance of an infrared Landau pole in the perturbative expression for the running coupling spawned the idea of infrared slavery, \emph{viz}.\ confinement expressed through a far-infrared divergence in the coupling.  In the absence of a nonperturbative definition of a unique running coupling, this idea is just a conjecture.  Notwithstanding that, and possibly inspired by the challenge, attempts to solve the confinement puzzle by completing the nonperturbative definition and calculation of a QCD running coupling have received ongoing attention, \emph{e.g}.\ Refs.\,\cite{Dokshitzer:1998nz, Grunberg:1982fw} and citations thereof.

The archetypal running coupling is that computed in QED \cite{GellMann:1954fq}, now known to great accuracy \cite{Tanabashi:2018oca}.  This Gell-Mann--Low effective charge is renormalisation group invariant (RGI) and process-independent (PI).  It is obtained by simply computing the photon vacuum polarisation because ghost fields decouple in Abelian theories.

Calculations in QCD are normally more difficult because ghost fields do not decouple.  However, combining the pinch technique (PT) \cite{Cornwall:1981zr, Cornwall:1989gv, Pilaftsis:1996fh, Binosi:2009qm} and background field method (BFM) \cite{Abbott:1980hw}, QCD can be made to ``look'' Abelian: one systematically rearranges classes of diagrams and their sums in order to obtain modified Schwinger functions that satisfy linear Slavnov-Taylor identities \cite{Taylor:1971ff, Slavnov:1972fg}.  In the gauge sector, using Landau gauge, this produces a modified gluon dressing function from which one can compute a unique QCD running coupling, which is RGI and PI.

\begin{figure}[t!]
\vspace*{1.5ex}

\includegraphics[width=0.95\linewidth]{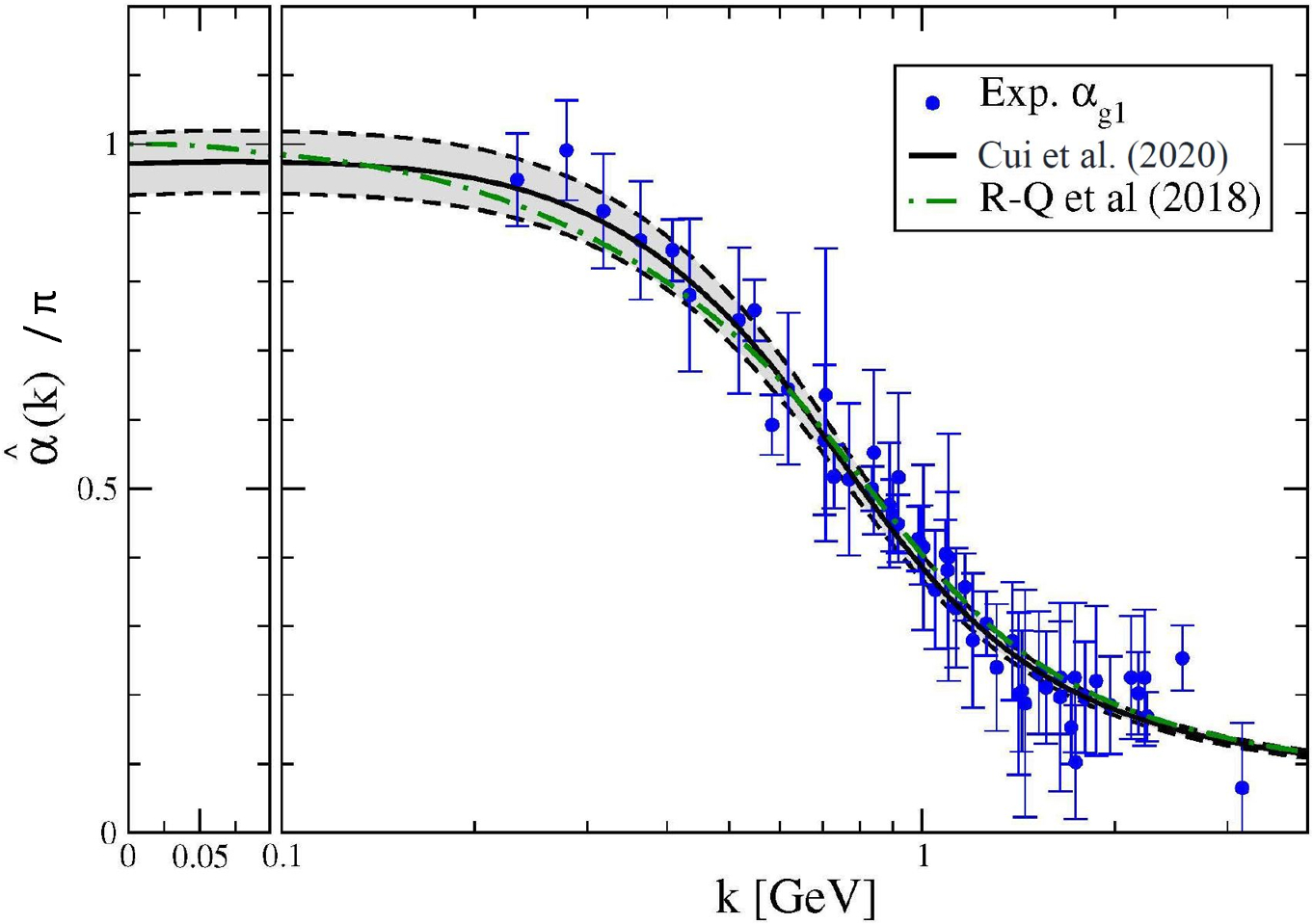}
\caption{\label{Figwidehatalpha}
Solid black curve within grey band -- RGI PI running-coupling, $\hat{\alpha}(k^2)/\pi$, computed in Ref.\,\cite{Cui:2019dwv} (Cui \emph{et al}.\ 2020); and dot-dashed green curve -- earlier result (R-Q \emph{et al}.\ 2018) \cite{Rodriguez-Quintero:2018wma}.
(The grey systematic uncertainty band bordered by dashed curves is explained in Ref.\,\cite{Cui:2019dwv}.)
For comparison, world data on the process-dependent charge, $\alpha_{g_1}$, defined via the Bjorken sum rule, are also depicted.  (The data sources are listed elsewhere \cite{Cui:2019dwv}.
For additional details, see Refs.\,\cite{Deur:2005cf, Deur:2008rf, Deur:2016tte}.)
The k-axis scale is linear to the left of the vertical partition and logarithmic otherwise.
}
\end{figure}

These notions lead to the effective coupling, $\hat\alpha(k^2)$, introduced in Ref.\,\cite{Binosi:2016nme}. Following improvements, particularly in the results from numerical simulations of lattice-regularised QCD (lQCD), the predictions have subsequently been refined \cite{Rodriguez-Quintero:2018wma, Cui:2019dwv}, with the result depicted in Fig.\,\ref{Figwidehatalpha}.  This charge is a unique strong-interaction analogue of the Gell-Mann--Low effective coupling in QED.  Owing to the dynamical breakdown of scale invariance, expressed through emergence of a RGI gluon mass-scale, with calculated value $m_0/{\rm GeV} = 0.43(1)$,
this running coupling saturates at infrared momenta:
\begin{equation}
\label{hatalpha0}
\frac{\hat\alpha(0)}{\pi}=0.97(4)\,.
\end{equation}
Importantly, the appearance of this mass-scale does not alter any Slavnov-Taylor identities \cite{Taylor:1971ff, Slavnov:1972fg}; hence, all aspects and consequences of QCD's BRST invariance \cite{Becchi:1975nq, Tyutin:1975qk} are preserved.

All properties of $\hat\alpha(k^2)$ are parameter-free predictions.  Moreover, this running coupling is \cite{Cui:2019dwv}:
(\emph{i}) pointwise (almost) identical to the process-dependent (PD) effective charge \cite{Dokshitzer:1998nz, Grunberg:1982fw}, $\alpha_{g_1}$, defined via the Bjorken sum rule -- as evident in Fig.\,\ref{Figwidehatalpha};
(\emph{ii}) capable of marking the boundary between soft and hard physics -- as discussed in Refs.\,\cite{Cui:2019dwv, Ding:2019qlr, Ding:2019lwe}; and
(\emph{iii}) that process-dependent charge which, used to integrate the one-loop DGLAP equations \cite{Dokshitzer:1977, Gribov:1972, Lipatov:1974qm, Altarelli:1977}, delivers agreement between pion parton distribution functions (PDFs) calculated at the hadronic scale and experiment \cite{Cui:2019dwv, Ding:2019qlr, Ding:2019lwe}.
In playing so many diverse r\^oles, $\hat\alpha(k^2)$ emerges as a strong candidate for that object which properly represents the interaction strength in QCD at any given momentum scale.

Examining Fig.\,\ref{Figwidehatalpha}, it is clear that the Landau pole, a prominent feature of perturbation theory, is screened (eliminated) in QCD.  This owes to the dynamical generation of a gluon mass-scale.  Hence, the theory possesses an infrared stable fixed point.\footnote{In support of this position, it is worth remarking that in the absence of an infrared stable fixed point the coupling would diverge at some infrared scale owing to the existence of a Landau pole.  No sign of such divergences is seen in experiment.  Moreover, lQCD also offers circumstantial support, \emph{viz}.\ there is no signal for the appearance of infrared divergences in the extrapolation of lattice results to the infinite-volume limit.}  Even more, it is evident in Fig.\,\ref{Figwidehatalpha} that QCD is approximately scale invariant on $0<k^2\lesssim m_0^2$ because the coupling practically stops running within this ``conformal window''.\footnote{Consequently, when used judiciously, a contact interaction can yield reliable estimates of integrated, infrared observables, such as masses and charges, \emph{e.g}.\ Refs.\,\cite{Roberts:2010rn, Roberts:2011wy, Chen:2012txa, Lu:2017cln, Raya:2017ggu, Yin:2019bxe}.}  In consequence of these features and with standard renormalisation theory ensuring the ultraviolet behaviour is under control, QCD emerges as a mathematically well-defined quantum field theory in four dimensions.

\section{Empirical Confinement}
\label{EmpConfinement}
Returning now to the question of confinement, Fig.\,\ref{Figwidehatalpha} establishes that typical potential-model pictures are unrealistic in assuming a ``confining potential'' between light quarks.  Instead, given that $(2 m_0/m_\pi)^2 \approx 40$, light-particle creation and annihilation effects are essentially nonperturbative in QCD; thus it is impossible in principle to compute a quantum mechanical potential between two light quarks \cite{Bali:2005fu, Prkacin:2005dc, Chang:2009ae}.

An alternative viewpoint associates confinement with dynamically-driven changes in the analytic structure of QCD's coloured propagators and vertices.  The existence of momentum-dependent gluon and quark masses ensures such outcomes.  Consequently, coloured $n$-point functions violate the axiom of reflection positivity, in which case the associated excitations are eliminated from the Hilbert space associated with asymptotic states \cite{GJ81}.

The violation of reflection positivity  is a sufficient condition for confinement \cite{Munczek:1983dx, Cahill:1985mh, Stingl:1985hx, Krein:1990sf, Burden:1991gd, Hawes:1993ef, Maris:1994ux, Roberts:1994dr, Bhagwat:2002tx, Roberts:2007ji, Bashir:2009fv, Strauss:2012dg, Bashir:2013zha, Qin:2013ufa, Lowdon:2015fig, Lucha:2016vte, Binosi:2016xxu, Binosi:2019ecz}.  It leads to a dynamical picture of this phenomenon.  Namely, suppose a gluon or quark is produced and begins to propagate.   After a short spacetime interval, on the order of $1/m_0\approx \tfrac{1}{2}\,$fm, an interaction occurs so that the parton loses its identity, sharing it with others.  This happens very often so that, finally, a cloud of partons is produced, which coalesces into hadron final states.  This is the physics of parton fragmentation functions (PFFs), which describe how QCD partons, generated in a high energy event and (nearly) massless in perturbation theory, convert into a shower of massive hadrons, \emph{i.e}.\ PFFs describe how hadrons with mass emerge from massless partons.

This perspective suggests that PFFs are the cleanest expression of dynamical confinement in QCD.  Moreover, PFFs are related to PDFs by crossing symmetry in the neighbourhood of their common boundary of support \cite{Gribov:1971zn}.  Hence, PFFs and PDFs can both provide fundamental insights into EHM.

The importance of PDFs and PFFs is further accentuated by the fact that, today, China and the USA are each separately working toward construction of high-luminosity electron-ion colliders, EicC and EIC, respectively.  Hadron tomography is a defining goal of both machines.  Namely, precision measurements that will enable science to draw three-dimensional images of the proton and other hadrons.  In so doing, they will chart the domain of confinement and its relationship to all other hadron properties.  At both facilities, the media of choice are generalised parton distributions (GPDs) and transverse momentum dependent parton distributions (TMDs).  Measurements of these quantities are expected to serve best in achieving hadron tomography.

Crucially, PDFs provide the boundary values for GPDs.  Moreover, every cross-section that can yield a given TMD involves a related PFF in convolution; so the PFF must be known precisely if the TMD is to be determined from the experiment.  However, despite their importance to the success of the roughly billion-dollar EicC and EIC projects, science is only just beginning to deliver sound predictions for pion PDFs; and there are no QCD-connected calculations of PFFs, GPDs and TMDs.  The difficulty is that although these distribution and fragmentation functions appear in cross-section formulae derived using perturbative QCD, they are essentially nonperturbative objects.  Hence, their reliable calculation requires use of nonperturbative methods; and such tools, whether continuum or lattice, still require further development.

\section{Pion as a Goldstone Mode}
\label{NGPion}
As the carriers of electric charge, quarks are visible in electron scattering experiments; hence, their distributions within hadrons can be measured.  In order to explain those measurements, it is necessary to know how quarks propagate.  This is described by the dressed-quark propagator:
\begin{subequations}
\label{Spgen}
\begin{align}
S(p)
 &= 1/[i \gamma\cdot p A(p^2) + B(p^2)] \\
& = Z(p^2)/[i\gamma\cdot p + M(p^2)]\,,
\end{align}
\end{subequations}
which is obtained as the solution of a gap equation whose kernel is critically dependent upon $\hat{\alpha}(k^2)$.  $M(p^2)$ in Eq.\,\eqref{Spgen} is the RGI dressed-quark mass-function.

The dynamical generation of a running gluon mass in QCD, characterised by the large mass-scale $m_0 \approx m_N/2$ at infrared momenta, transmits into the matter sector with the analogous emergence of $M(k^2)$ \cite{Bhagwat:2003vw, Bowman:2005vx, Bhagwat:2006tu}.  Notably, $M(0) \approx m_N/3$ in the chiral limit.  This mass-function is nonzero even in the absence of a Higgs mechanism, \emph{viz}.\ it is a signature feature of DCSB \cite{Nambu:2011zz}.

DCSB is a pivotal emergent phenomenon in QCD.  It is expressed in hadron wave functions \cite{Brodsky:2009zd, Brodsky:2010xf, Chang:2011mu, Brodsky:2012ku, Roberts:2015lja}; and in serving as the physical origin of a constituent-like quark mass-scale, DCSB may be viewed as the source for more than 98\% of the visible mass in the Universe.  Moreover, since classical massless chromodynamics is a scale-invariant theory,  it follows that DCSB is fundamentally connected with the \emph{origin of mass from nothing}.

It is important to insist that chiral symmetry breaking in the absence of a Higgs mechanism is \underline{dynamical}, as distinct from spontaneous, because nothing is added to QCD in order to effect this outcome and there is no simple change of variables in the QCD action that will make it apparent.  Instead, through the act of quantising the classical chromodynamics of massless gluons and quarks, a large mass-scale is generated in both the gauge and matter sectors.

DCSB is empirically revealed very clearly in properties of the pion, whose structure in QCD is described by a Bethe-Salpeter amplitude:
\begin{align}
\nonumber
\Gamma_{\pi}(k;P) & = \gamma_5 \left[
i E_{\pi}(k;P) + \gamma\cdot P F_{\pi}(k;P)  \right.\\
& \quad \left. +\, \gamma\cdot k \, G_{\pi}(k;P) + \sigma_{\mu\nu} k_\mu P_\nu H_{\pi}(k;P) \right],
\label{genGpi}
\end{align}
where $k$ is the relative momentum between the  valence-quark and -antiquark constituents (defined here such that the scalar functions in Eq.\,\eqref{genGpi} are even under $k\cdot P \to - k\cdot P$) and $P$ is their total momentum.  $\Gamma_{\pi}(k;P)$ is simply related to an object that would be the pion's Schr\"odinger wave function if a nonrelativistic limit were appropriate.

In chiral-limit QCD, if, and only if, chiral symmetry is dynamically broken, then \cite{Maris:1997hd, Qin:2014vya, Binosi:2016rxz}:
\begin{equation}
\label{gtrE}
f_\pi^0 E_\pi(k;0) = B(k^2)\,,
\end{equation}
where $f_\pi^0$ is the chiral-limit value of the pion's leptonic decay constant.
This identity is remarkable.  It is true in any covariant gauge and independent of the renormalisation scheme; and it means that the two-body problem is solved, nearly completely, once the solution to the one body problem is known.

Eq.\,\eqref{gtrE} is the most basic statement in QCD of the Nambu-Goldstone theorem \cite{Nambu:1960tm, Goldstone:1961eq}.  It entails that pion properties are an almost direct measure of the dressed-quark mass function.  Thus, enigmatically, the qualities of the nearly-massless pion are the cleanest expression of the mechanism that is responsible for virtually all visible mass in the Universe.

As explained elsewhere \cite{Roberts:2016vyn}, Eq.\,\eqref{gtrE} is the keystone of a proof that the pion remains massless in the chiral limit irrespective of the emergence of a large gluon mass scale which drives the nucleon mass to 1\,GeV.  In fact, in the pseudoscalar channel, the sum of the dynamically generated masses of the quark and antiquark is precisely cancelled by the attractive interaction energy between these dressed constituents if, and only if, Eq.\,\eqref{gtrE} is preserved:
\begin{align}
\nonumber
& M^{\rm dressed}_{\rm quark} + M^{\rm dressed}_{\rm antiquark} \\
& + U^{\rm dressed}_{\rm quark-antiquark\;interaction} \stackrel{\rm chiral\;limit}{\equiv} 0\,.
\label{EasyOne}
\end{align}
This guarantees the \emph{disappearance} of the scale anomaly in the chiral-limit pion.  Eq.\,\eqref{EasyOne} is not merely ``hand-waving''.  Rather, it sketches the cancellations that take place in the pseudoscalar projection of the fully-dressed quark+antiquark scattering matrix, which can be displayed rigorously \cite{Bender:1996bb}.

It has been known for more than fifty years that if one reintroduces the Higgs mechanism, so that light-quark current masses appear in the QCD Lagrangian, then \cite{GellMann:1968rz}:
\begin{align}
\label{GMORO}
m_\pi^2 & = (m_u^\zeta + m_d^\zeta) \frac{\kappa^\zeta}{f_\pi^2}\,,
\end{align}
where $\kappa^\zeta$ is the in-pion chiral condensate \cite{Brodsky:2010xf, Chang:2011mu, Brodsky:2012ku, Roberts:2015lja}.  ($\zeta$ is the renormalisation scale.)  Eq.\,\eqref{GMORO} is a Poincar\'e-invariant decomposition of the pion mass.  It states that the entirety of the pion-mass-squared is generated by the Lagrangian current-quark mass term; and, moreover, that the sum of current-quark masses is multiplied by an EHM enhancement factor, because both $\kappa^\zeta$, $f_\pi$ are order parameters for DCSB \cite{Bender:1996bm}.  (Naturally, $\kappa^\zeta$, $f_\pi$ also depend on $(m_u^\zeta + m_d^\zeta)$, but such higher order effects are quantitatively small and qualitatively immaterial in the light-quark sector \cite{Maris:1997tm}.  Analogous statements are also true for the examples that follow in this section.)

One can reveal the strength of the DCSB magnifier by rewriting Eq.\,\eqref{GMORO} in the following form:
\begin{subequations}
\begin{align}
\label{GMORQM}
m_\pi & = (m_u^\zeta + m_d^\zeta)\, {\mathpzc s}_\zeta \,, \\
{\mathpzc s}_\zeta^2 & = \frac{\kappa^\zeta}{ (m_u^\zeta + m_d^\zeta) f_\pi^2}.
\end{align}
\end{subequations}
In bound-states that are readily described by relativistic quantum mechanics one would have ${\mathpzc s}\approx 1$, \emph{viz}.\
\begin{align}
\nonumber &
{\rm system\,mass}& \\
& \approx \sum_{i\,{\rm in\,system}} ({\rm Lagrangian\,mass\,of\,constituent})_i \,.
\label{RelativisticMassEquation}
\end{align}
Here, however, a description of the physical pion requires ($\zeta_2=2\,$GeV) \cite{Tanabashi:2018oca}:
\begin{equation}
{\mathpzc s}_{\zeta_2} = 20\,,
\end{equation}
which is a very large EHM-induced enhancement factor.  In the context of Eq.\,\eqref{EasyOne}, this outcome reveals just how neat is the balance between one-body dressing and two-body interaction effects in the pion bound-state: disturb the pion composite with just a small amount of Higgs-generated mass and the bound-state's mass rises very quickly in response.

The pion's valence-quark spin-flip partner is the $\rho$-meson, for which one can write \cite{Flambaum:2005kc}:
\begin{align}
m_\rho = m_\rho^0 +   (m_u^{\zeta_2} + m_d^{\zeta_2}) \, h_\rho^{\zeta_2}\,,
\end{align}
where the chiral-limit $\rho$-mass, $m_\rho^0 \approx 0.75\,$GeV$\approx 2 M(0)$, owes entirely to EHM.  (Obviously, $m_\pi^0 \equiv 0$.)  Here, the current-mass enhancement factor $h_\rho^{\zeta_2} \approx 3.6$, roughly 3.6-times the value for a typical system in quantum mechanics.  The size of $h_\rho^{\zeta_2}$, too, owes largely to EHM.

A similar analysis applies for the proton \cite{Flambaum:2005kc}:
\begin{align}
m_p = m_p^0 +   (2 m_u^{\zeta_2} + m_d^{\zeta_2}) \, h_p^{\zeta_2}\,;
\end{align}
once again, $m_p^0 \approx 3 M(0)$ owes entirely to EHM; which is also the primary source for the magnitude of the current-mass enhancement factor, $h_p^{\zeta_2} \approx 7.1$.

These three examples illustrate some general rules for hadron masses:
(\emph{i}) estimates based on notions familiar from relativistic quantum mechanics, Eq.\,\eqref{RelativisticMassEquation}, typically arrive at only $\sim 1$\% of a hadron's mass -- hence, the Higgs mechanism alone is responsible for just $\sim 1$\% of visible mass;
(\emph{ii}) the contribution of the current-mass term in QCD's Lagrangian is strongly enhanced as a consequence of EHM, in particular for the pion;
and (\emph{iii}) in all systems for which no symmetry ensures Eq.\,\eqref{EasyOne}, EHM is key to more than 98\% of a hadron's mass.
It is worth highlighting that these features lay the foundation which guarantees, \emph{inter alia}, accuracy of equal spacing rules in the hadron spectrum \cite{Okubo:1961jc, GellMann:1962xb, Qin:2018dqp, Qin:2019hgk}.

\section{Expressions of EHM in NG Mode Parton Distributions}
\label{ObservableEHM}
QCD's interactions are universal, \emph{i.e}.\ they are the same in all hadrons.  Hence, cancellations similar to those indicated by Eq.\,\eqref{EasyOne} take place within the proton.  However, as just noted, in the proton, no symmetry requires the cancellations to be complete.  Thus, the value of the proton's mass is typical of the magnitude of scale breaking in the one-body sectors, \emph{viz}.\ the gluon and quark mass scales, $m_0$ and $M(0)$, respectively.  In fact, no significant hadronic mass scale is possible unless one of similar size is expressed in the dressed-propagators of gluons and quarks.  It follows that the mechanism(s) responsible for EHM can be exposed by measurements sensitive to such dressing.  This potential is offered by a large array of observables, \emph{e.g}.: spectra and static properties; form factors -- elastic and transition; and all types of parton distributions.
Hereafter, we sketch a few topical examples.

\begin{figure}[t]
\vspace*{-3ex}

\includegraphics[width=0.46\textwidth]{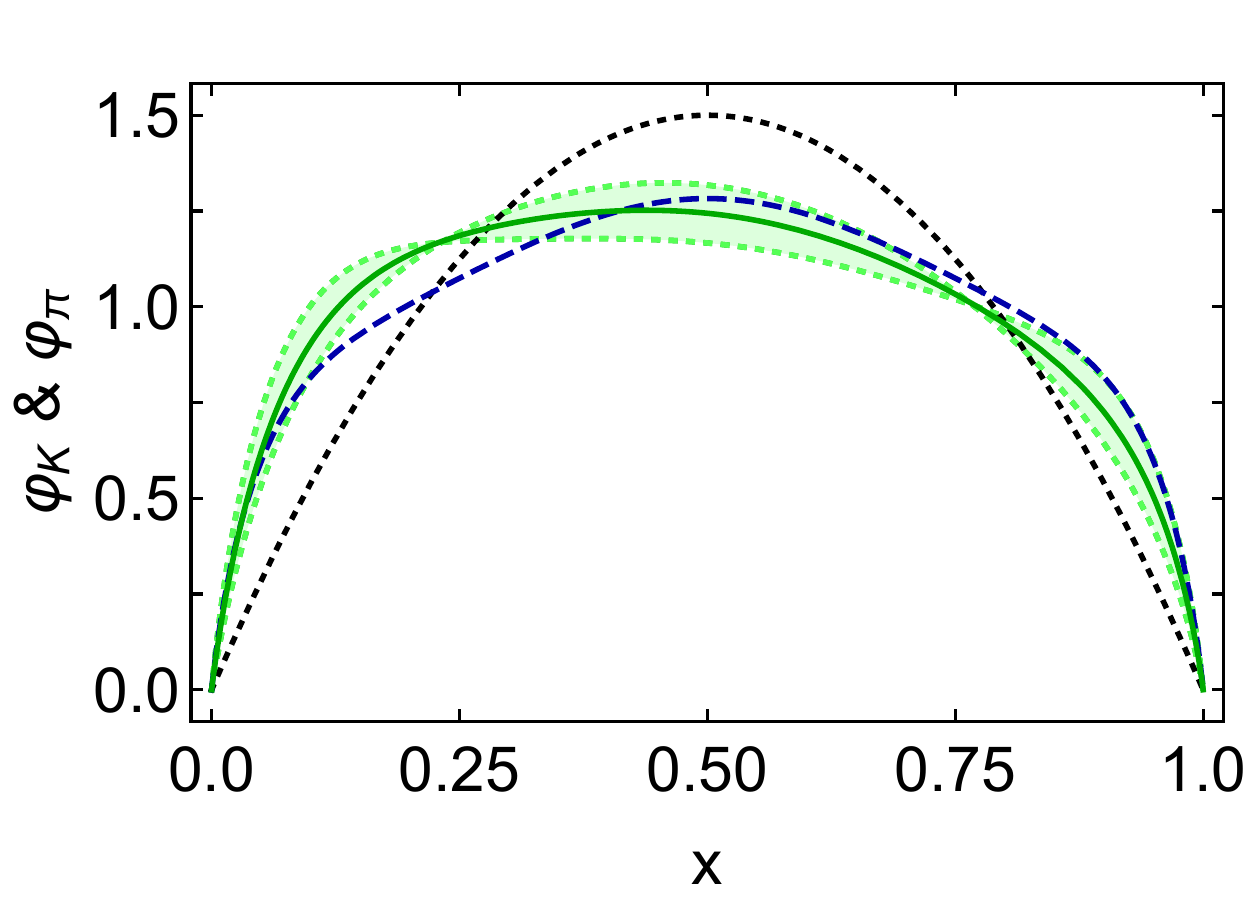}
\caption{\label{FigPDAs}
Parton distribution amplitudes (PDAs) for the $\pi$ meson (dashed blue curve) -- based on the results in Refs.\,\cite{Chang:2013pq, Segovia:2013eca}; and kaon (solid green curve within the shaded green band -- based on the discussion in Ref.\,\cite{Horn:2016rip}.
(The PDAs are drawn at the hadronic scale, $\zeta_H=m_0$, whose value is explained elsewhere \cite{Ding:2019qlr, Ding:2019lwe, Cui:2019dwv}.)
The asymptotic profile \cite{Lepage:1979zb, Efremov:1979qk, Lepage:1980fj}: $\varphi_{\rm as}(x)=6 x(1-x)$, is drawn as the dotted black curve.
Following intense effort, the pion PDA is well known \cite{Brodsky:2006uqa, Chang:2013pq, Segovia:2013eca}; consequently, we only display uncertainty on the kaon PDA,  the origin of which is discussed in Ref.\,\cite[Sec.\,6.2]{Horn:2016rip}.
Evidently, all PDAs are broadened with respect to $\varphi_{\rm as}(x)$; and the peak in the kaon's distribution amplitude is shifted from $x_\pi=0.5$ to $x_K=0.43(4)$.  Both effects are consequences of EHM.  Notably, $x_\pi/x_K \approx f_K/f_\pi$, \emph{viz}.\ the ratio of meson leptonic decay constants, each of which is an order parameter for DCSB.  This relationship between the magnitude of SU$(3)$-flavour symmetry breaking and the mass-scales associated with EHM is observed in all relevant observables.  (See, \emph{e.g}.\ Refs.\,\cite{Segovia:2013eca, Shi:2015esa, Chen:2016sno, Gao:2017mmp}.)
}
\end{figure}

\subsection{Pion Parton Distributions}
Considering only their valence-quark content, pions are Nature's simplest hadrons: $\pi^+  \sim u\bar d$, $\pi^- \sim d \bar u$, $\pi^0 \sim u\bar u - d\bar d$.  Hence, a basic quantity in any discussion of their structure is the associated distribution function, ${\mathpzc q}^\pi(x;\zeta)$.  This density charts the probability that a valence ${\mathpzc q}$-quark in the pion carries a light-front fraction $x$ of the system's total momentum; and one of the earliest predictions of the parton model, augmented by features of perturbative QCD (pQCD), is \cite{Ezawa:1974wm, Farrar:1975yb, Berger:1979du, Chang:2013pq, Chang:2020kjj}:
\begin{equation}
\label{PDFQCD}
{\mathpzc q}^{\pi}(x\simeq 1;\zeta =\zeta_H) \propto (1-x)^{2}\,,
\end{equation}
where $\zeta_H$ is an energy scale characteristic of strong gauge-sector dynamics \cite{Ding:2019qlr, Ding:2019lwe, Cui:2019dwv}.  Moreover, the exponent evolves as $\zeta$ increases beyond $\zeta_H$, becoming $2+\gamma$, where $\gamma\gtrsim 0$ is an anomalous dimension that increases logarithmically with $\zeta$.  (Using  ${\mathpzc G}$-parity symmetry, a good approximation in the Standard Model, $\bar d^{\pi^+}(x)=u^{\pi^+}(x) $.)

${\mathpzc q}^\pi(x)$ is measurable in $\pi$-nucleon Drell-Yan (DY) experiments \cite{Badier:1980jq, Badier:1983mj, Betev:1985pg, Falciano:1986wk, Guanziroli:1987rp, Conway:1989fs, Heinrich:1991zm}. However, conclusions drawn from analyses of these experiments are controversial \cite{Holt:2010vj}.
For instance, using a leading-order (LO) pQCD analysis of their data, Ref.\,\cite{Conway:1989fs} (the E615 experiment) reported 
\begin{align}
\label{E615result}
{\mathpzc q}_{\rm E615}^{\pi}(x\simeq 1; \zeta_5=5.2\,{\rm GeV}) \propto  (1-x)^{1}\,,
\end{align}
in conflict with Eq.\,\eqref{PDFQCD}.  Subsequent calculations \cite{Hecht:2000xa} confirmed Eq.\,\eqref{PDFQCD}, prompting reconsideration of the E615 analysis, with the result that, at next-to-leading order (NLO) and including soft-gluon resummation \cite{Wijesooriya:2005ir, Aicher:2010cb}, the E615 data become consistent with Eq.\,\eqref{PDFQCD}.

\begin{figure}[t]
\includegraphics[clip, width=0.44\textwidth]{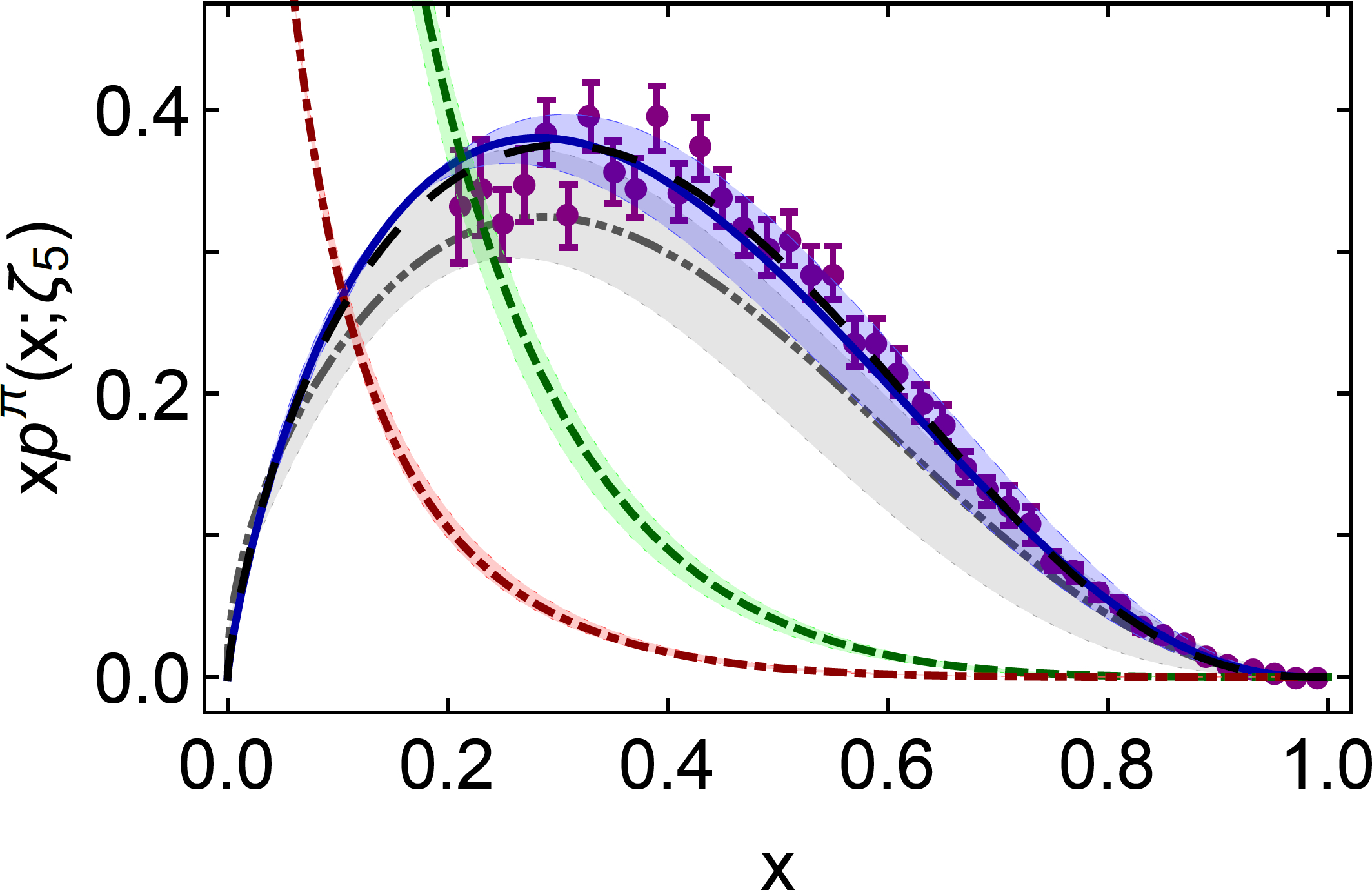}
\caption{\label{figF12}
Pion valence-quark momentum distribution function, $x {\mathpzc u}^\pi(x;\zeta_5)$:
solid blue curve  -- modern continuum calculation \cite{Ding:2019qlr, Ding:2019lwe};
long-dashed black curve -- early continuum analysis \cite{Hecht:2000xa};
and dot-dot-dashed grey curve -- lQCD result \cite{Sufian:2019bol}.
Gluon momentum distribution, $x g^\pi(x;\zeta_5)$ -- dashed green curve;
and sea-quark momentum distribution, $x S^\pi(x;\zeta_5)$ -- dot-dashed red curve.
Data [purple] from Ref.\,\cite{Conway:1989fs}, rescaled following Ref.\,\cite{Aicher:2010cb}.
(The shaded bands indicate the size of calculation-specific uncertainties \cite{Ding:2019qlr, Ding:2019lwe}.  Our convention in drawing such plots is $\langle x(2 q^\pi +S^\pi +g^\pi )\rangle =1$.)
}
\end{figure}

Notwithstanding this progress, uncertainty over Eq.\,\eqref{PDFQCD} will remain until other analyses of the E615 data incorporate threshold resummation effects; a step whose importance is acknowledged \cite{Barry:2018ort}: ``\emph{Furthermore, the present analysis does not include threshold resummation effects, which are known to be important at large} $x$ \cite{Aicher:2010cb, Westmark:2017uig}, \emph{and this will be examined in a separate analysis}.''  Independent of that, modern data must be obtained.  The latter are forthcoming because relevant tagged deep-inelastic scattering experiments are approved at Jefferson Laboratory (JLab) \cite{Keppel:2015, Keppel:2015B, McKenney:2015xis} and the goal has high priority at other existing and anticipated facilities \cite{Petrov:2011pg, Peng:2016ebs, Peng:2017ddf, Horn:2018fqr, Denisov:2018unj, Aguilar:2019teb, EicCWP}.
%
Meanwhile, models that are compatible with Eq.\,\eqref{E615result} but inconsistent with Eq.\,\eqref{PDFQCD}, \emph{e.g}.\ Refs.\,\cite{Broniowski:2007si, deTeramond:2018ecg, Lan:2019rba}, continue to draw support from careful but incomplete analyses of data \cite{Chang:2020kjj}.

Fortunately, real progress in QCD theory continues unabated.  For instance, novel lQCD algorithms \cite{Liu:1993cv, Ji:2013dva, Radyushkin:2016hsy, Radyushkin:2017cyf, Chambers:2017dov} are beginning to yield results for the pointwise behaviour of the pion's valence-quark distribution \cite{Xu:2018eii, Chen:2018fwa, Oehm:2018jvm, Karthik:2018wmj, Sufian:2019bol}.

Furthermore, extensions of the continuum analysis in Ref.\,\cite{Hecht:2000xa} have yielded the first parameter-free predictions of the valence, glue and sea distributions within the pion \cite{Ding:2019qlr, Ding:2019lwe}; and revealed that, like the pion's leading-twist parton distribution amplitude (PDA), depicted in Fig.\,\ref{FigPDAs}, the valence-quark distribution function is hardened by DCSB, \emph{i.e}.\ as an immediate consequence of EHM.

The continuum predictions from Refs.\,\cite{Ding:2019qlr, Ding:2019lwe} are depicted in Fig.\,\ref{figF12}.  Portentously, the result for ${\mathpzc u}^\pi(x;\zeta_5)$, \emph{i.e}.\ the solid blue curve in Fig.\,\ref{figF12}, matches that obtained using lQCD \cite{Sufian:2019bol}.  (Here, we have included the theory uncertainty band described in Refs.\,\cite{Ding:2019qlr, Ding:2019lwe}, which reflects the precision in $\hat\alpha(k^2=0)$, Eq.\,\eqref{hatalpha0}.  Note, too, that the lQCD result is consistent with the prediction in Ref.\,\cite{Hecht:2000xa}, made twenty years ago.)  Evidently, a modern confluence has been reached: two disparate treatments of the pion bound-state problem have arrived at the \emph{same} prediction for the pion's valence-quark distribution function, thus demonstrating that real strides are being made toward understanding pion structure.   Plainly, the Standard Model prediction, Eq.\,\eqref{PDFQCD}, is stronger than ever before; and an era is dawning in which the ultimate experimental checks can be made \cite{R.A.Montgomery:2017hab, Denisov:2018unj, Aguilar:2019teb}.

\begin{figure}[t]
\vspace*{-3.5ex}

\includegraphics[width=0.44\textwidth]{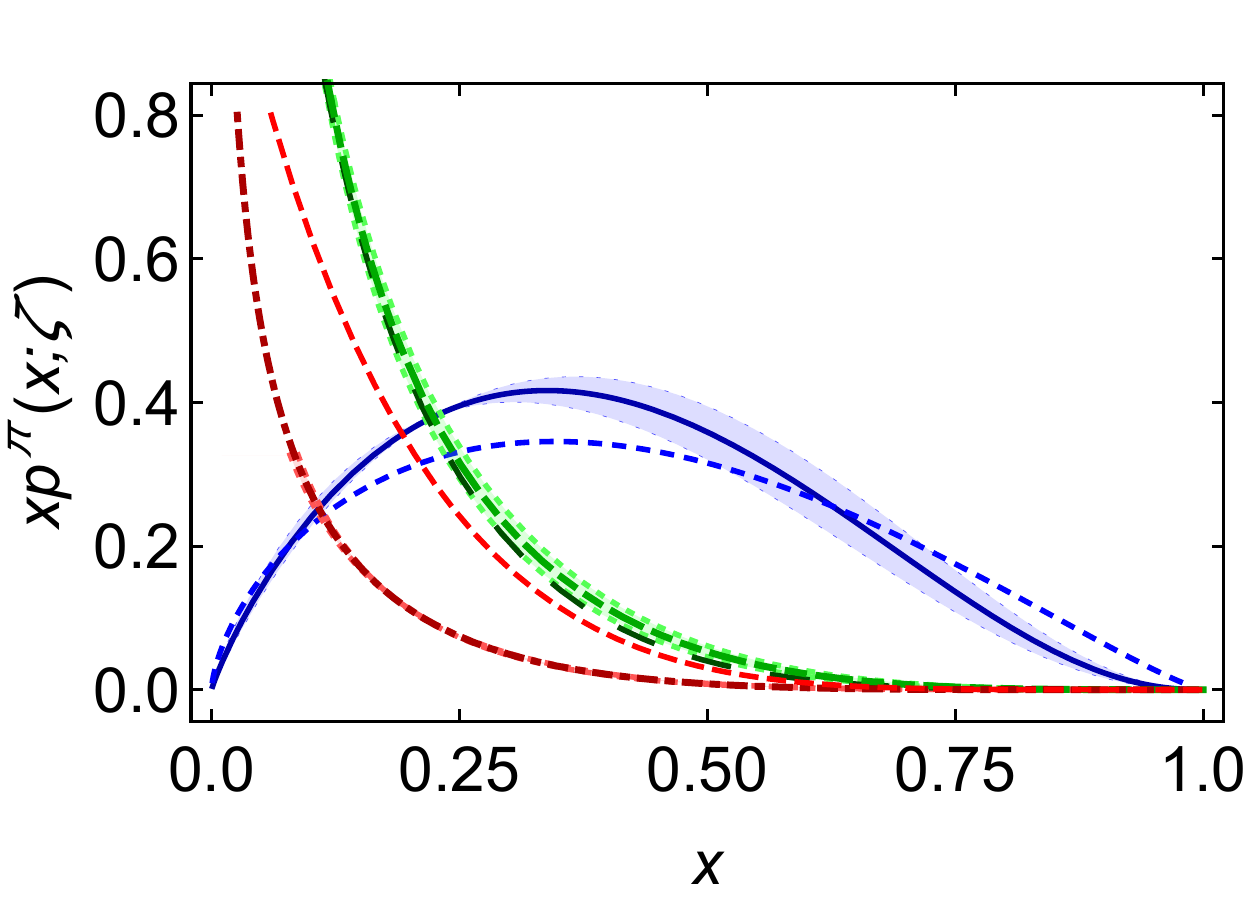}
\caption{\label{FigcfJAM} Distributions from Refs.\,\cite{Ding:2019qlr, Ding:2019lwe} evaluated at $\zeta=\zeta_2=2\,$GeV:
$p=\,$valence -- solid blue curve;
$p=\,$glue -- long dot-dot-dashed green;
and $p=\,$sea -- dot-dashed red.
Phenomenological results ($\zeta = 3.2\,$GeV) from Ref.\,\cite[Fig.\,2]{Barry:2018ort} are plotted for comparison: $p=\,$valence -- short-dashed blue;
$p=\,$glue -- long-dashed dark-green; and
$p=\,$sea -- dashed red.
}
\end{figure}

Fig.\,\ref{figF12} also displays predictions for the sea and glue distributions within the pion.  Context for those results is provided in Fig.\,\ref{FigcfJAM}, which compares the predictions from Refs.\,\cite{Ding:2019qlr, Ding:2019lwe} with fits from a global analysis of combined $\pi N$ DY and leading-neutron electroproduction data \cite{Chekanov:2002pf, Aaron:2010ab}.  DGLAP evolution is logarithmic and the $\zeta$-scales are similar; hence, direct comparisons are valid.

Regarding Fig.\,\ref{FigcfJAM}, the following remarks are required.
(\emph{i}) Even though both depicted valence distributions yield compatible momentum fractions ($\langle 2 x\rangle^\pi =0.48(3)$ \cite{Ding:2019qlr, Ding:2019lwe} \emph{cf}.\ $\langle 2 x\rangle^\pi =0.49(2)$ \cite{Barry:2018ort}, the $x$-profiles are very different.  This is readily understood.  The results in Refs.\,\cite{Ding:2019qlr, Ding:2019lwe} are predictions obtained using a symmetry-preserving truncation of QCD's continuum bound-state equations.  They preserve QCD constraints on PDFs and agree with the complete next-to-leading-order analysis of E615 data in Ref.\,\cite{Aicher:2010cb}, which includes threshold resummation effects.  On the other hand, the analysis in Ref.\,\cite{Barry:2018ort} ignores threshold resummation and delivers a result for the valence distribution that is inconsistent with QCD, as it manifests in Eq.\,\eqref{PDFQCD}.  The future here is improved phenomenological analyses and a series of new experiments.

(\emph{ii}) The gluon distribution predicted in Refs.\,\cite{Ding:2019qlr, Ding:2019lwe} and that fitted in Ref.\,\cite{Barry:2018ort}: yield compatible momentum fractions, $0.41(2)$ \emph{cf}.\ $0.35(3)$; and are pointwise similar on $x>0.05$.  This is notable because the glue distributions in Fig.\,\ref{FigcfJAM} are quite different from those inferred in earlier analyses \cite{Gluck:1999xe, Sutton:1991ay}; an outcome which highlights the need for new experiments that are directly sensitive to the pion's gluon content.  Such a need can be addressed by measurements of prompt photon and $J/\Psi$ production.  Here, new experiments are proposed at CERN by the COMPASS++/AMBER Collaboration \cite{Denisov:2018unj}.

(\emph{iii}) Refs.\,\cite{Ding:2019qlr, Ding:2019lwe} predict $\langle x\rangle_{\rm sea}^\pi =0.11(2)$ \emph{cf}.\ $\langle x\rangle_{\rm sea}^\pi =0.16(2)$ in Ref.\,\cite{Barry:2018ort}.  This $\sim 35$\% difference is expressed in markedly different $x$-profiles for the sea-quark distributions plotted in Fig.\,\ref{FigcfJAM}.  Thus if the pion's gluon content is considered uncertain, then it would be fair to describe the sea-quark distribution as empirically unknown.  Hence, there is excellent cause to follow the suggestion in Ref.\,\cite{Londergan:1995wp} and seek information by collecting DY data using $\pi^\pm$ beams on isoscalar targets \cite{Denisov:2018unj}.

It is worth highlighting that EHM-induced broadening of the pion PDF as found in Refs.\,\cite{Ding:2019qlr, Ding:2019lwe} is crucial to the agreement with data \cite{Aicher:2010cb} and the new lQCD result \cite{Sufian:2019bol}.  With hindsight, the existence of such broadening is not surprising.  It could have been anticipated from the relationship existing between leading-twist PDAs and valence-quark PDFs, expressed via a meson's light-front wave function, $\psi(x,k_\perp^2)$, where $k_\perp$ is the light-front transverse momentum:
{\allowdisplaybreaks
\begin{subequations}
\begin{align}
\varphi(x) & \sim \int d^2 k_\perp \psi(x,k_\perp^2)\,, \\
{\mathpzc q}(x) & \sim \int d^2 k_\perp |\psi(x,k_\perp^2) |^2\,.
\end{align}
\end{subequations}}
\hspace*{-0.3\parindent}Indeed, since factorised representations of pion and kaon light-front wave functions are often reliable for integrated quantities \cite{Xu:2018eii}, it is a good approximation to write
\begin{equation}
\label{PDFeqPDA2}
{\mathpzc q}_{\pi,K}(x;\zeta_H) \propto \varphi^{\mathpzc q}_{\pi,K}(x;\zeta_H)^2,
\end{equation}
where the constant of proportionality is fixed by baryon number conservation.  Owing to parton splitting effects, Eq.\,\eqref{PDFeqPDA2} is not valid on $\zeta>\zeta_H$.
Nevertheless, since the evolution equations for both PDFs and PDAs are known, the connection is not lost; it merely metamorphoses.

\begin{figure}[t]
\vspace*{-2.5ex}

\includegraphics[width=0.44\textwidth]{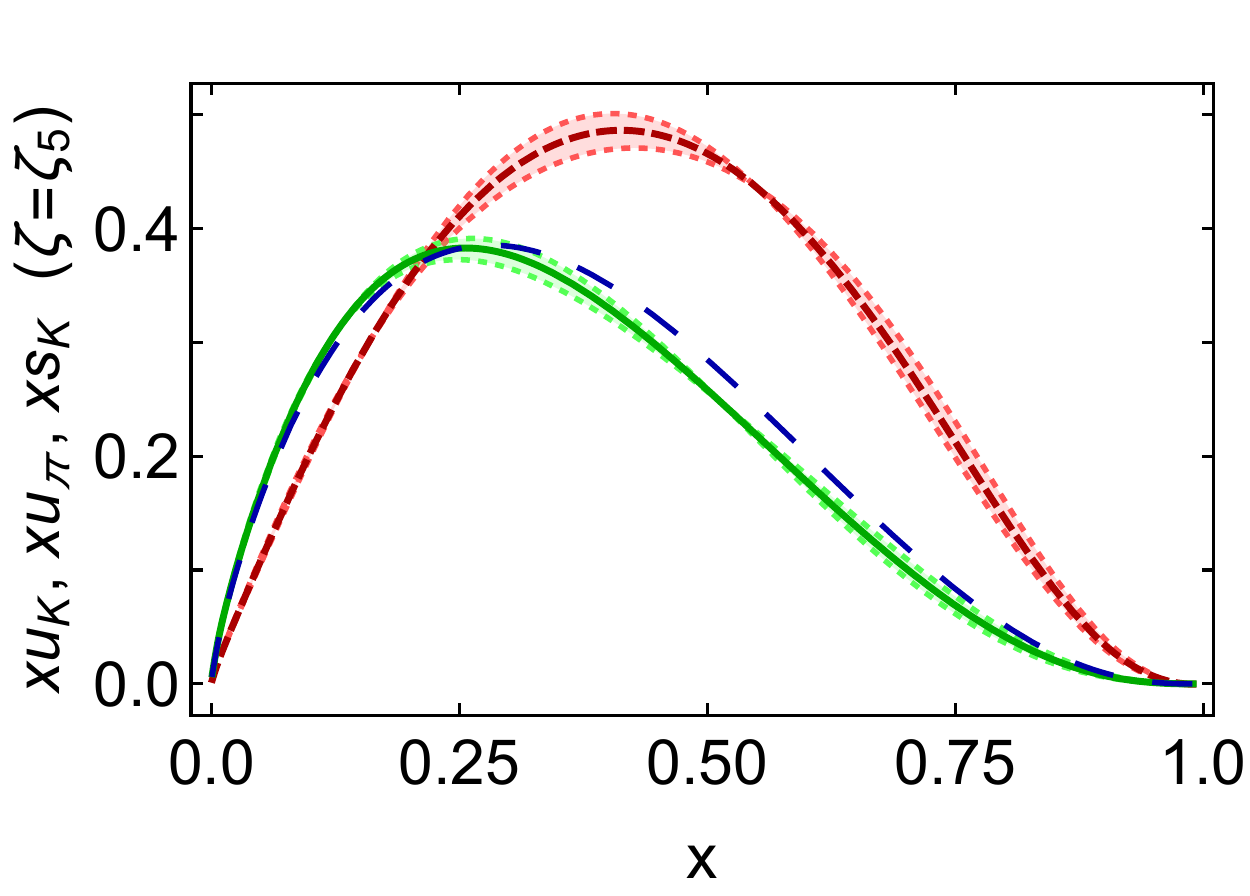}
\caption{\label{FigPDFs} Parton distribution functions corresponding to the PDAs in Fig.\,\ref{FigPDAs}: ${\mathpzc u}^{\pi}(x;\zeta_5)$  -- dashed blue curve; and ${\mathpzc u}^K(x;\zeta_5)$ -- solid green curve within green bands.
The dashed red curve within red bands is $\bar{\mathpzc s}^K(x;\zeta_5)$, computed as described in connection with Eq.\,\eqref{MassDependent}.}
\end{figure}

Employing Eq.\,\eqref{PDFeqPDA2}, one obtains the $\pi$- and $K$-meson ${\mathpzc u}$-quark distribution functions depicted in Fig.\,\ref{FigPDFs} when using the approximation to $\hat\alpha(k^2)$ detailed in Ref.\,\cite[Eq.\,(23)]{Ding:2019lwe} to integrate the one-loop DGLAP equations \cite{Dokshitzer:1977, Gribov:1972, Lipatov:1974qm, Altarelli:1977} and evolve ${\mathpzc u}^{\pi,K}(x;\zeta_H)$ to ${\mathpzc u}^{\pi,K}(x,\zeta_5)$.  (As a first step, we used splitting kernels for massless partons.  Given the importance of EHM, this should be improved in future because, \emph{e.g}.\ heavier quarks radiate gluons less readily than lighter quarks and gluons produce fewer heavy quark+antiquark pairs \cite{Landau:1953um, Migdal:1956tc}.  Such effects are ignored in massless splitting functions \cite{Catani:1996sc, Kim:2001kh, Pascaud:2011zt}.)

\subsection{Balancing EHM and the Higgs Mechanism}
\begin{figure}[t]
\includegraphics[clip, width=0.44\textwidth]{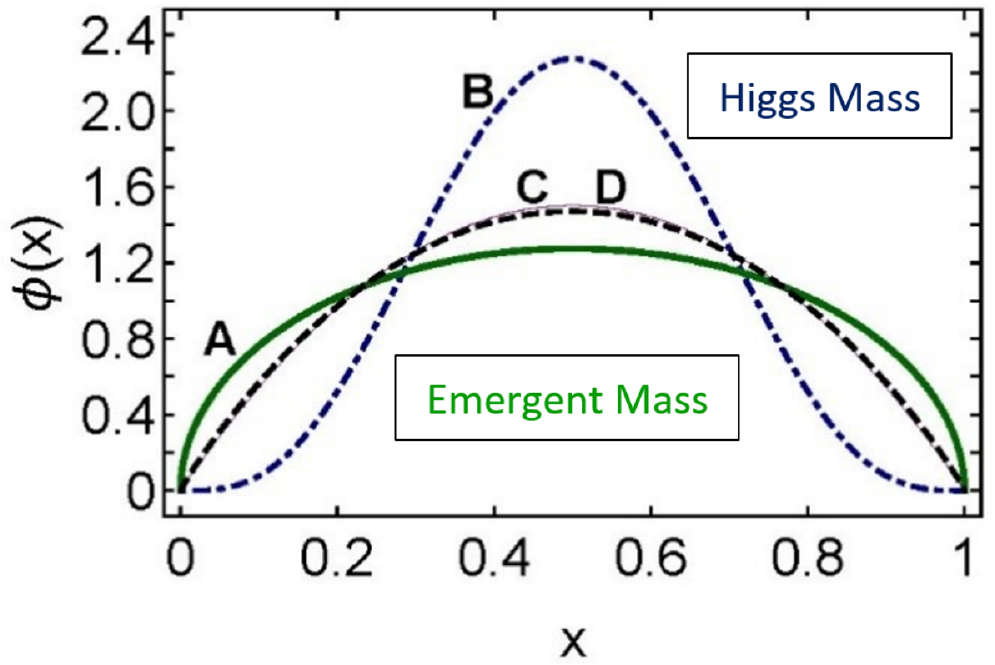}
\caption{\label{figPDAsO}
Twist-two parton distribution amplitudes at a resolving scale $\zeta=2 \,$GeV$=:\zeta_2$. \textbf{A} solid (green) curve – pion $\Leftarrow$ emergent mass generation is dominant; \textbf{B} dot-dashed (blue) curve – $\eta_c$ meson $\Leftarrow$ Higgs mechanism is the primary source of mass generation;  \textbf{C} solid (thin, purple) curve -- asymptotic profile, $\varphi_{\rm as}(x)$; and \textbf{D} dashed (black) curve – ``heavy-pion'', \emph{i.e}.\ a pion-like pseudo-scalar meson in which the valence-quark current masses take values corresponding to a strange quark $\Leftarrow$ the boundary, where emergent and Higgs-driven mass generation are equally important.
}
\end{figure}

The potential of pion structure function measurements to expose emergent mass is greatly enhanced if one includes similar kaon measurements.  This is illustrated by considering leading-twist meson PDAs, a number of which are depicted in Fig.\,\ref{figPDAsO}.  The image answers the following question: When does the Higgs mechanism begin to influence mass generation?  In the limit of infinitely-heavy quarks, \emph{i.e}.\ when the Higgs mechanism has overwhelmed every other mass generating force, the PDA becomes $\delta(x-1/2)$. The heavy $\eta_c$ meson, constituted from a valence charm-quark and its antimatter partner, feels the Higgs mechanism strongly.  Conversely, contemporary calculations predict that, compared with the scale-free asymptotic profile, $\varphi_{\rm as}(x)$, the PDA for the light-quark pion is a broadened, flattened, unimodal function \cite{Chang:2013pq, Zhang:2017bzy}. Such features are a definitive signal that pion properties express emergent mass generation. The remaining example in Fig.\,\ref{figPDAsO} is the PDA for a system composed of strange quarks.  It almost matches $\varphi_{\rm as}(x)$; thus, this system lies in the neighbourhood of the boundary, with strong (emergent) mass generation and the weak (Higgs-connected) mass playing a roughly equal r\^ole.

It follows that comparisons between distributions within systems constituted solely from light valence quarks and those associated with systems containing strange quarks are ideally suited to exposing measurable signals of EHM in counterpoint to Higgs-driven effects.  An example may be found in the contrast between the valence-quark PDFs of the pion and kaon at large Bjorken-$x$.  Here, a significant disparity between the distributions could point to a marked difference between the fractions of pion and kaon momentum carried by the other bound state participants, particularly gluons.

\subsection{Host-dependence of light-quark distributions}
A prediction for the ratio ${\mathpzc u}^K(x;\zeta_5)/{\mathpzc u}^\pi(x;\zeta_5)$ was delivered in \cite{Chen:2016sno}.  The results are consistent with the viewpoint presented above, \emph{viz}.\ the gluon content of the kaon at $\zeta_K=0.51\,$GeV is just $5\pm 5$\%, whereas that for the pion is $\sim 30$\%.  Ref.\,\cite{Chen:2016sno} found that these marked differences between the gluon content of the pion and kaon persist to large resolving scales, \emph{e.g}.\ at $\zeta = 2\,$GeV, the gluon momentum fraction in the pion is still $\sim 40$\% greater than that in the kaon.  Moreover, this difference in gluon content is expressed in the large-$x$ behaviour of the $\pi$ and $K$ valence-quark PDFs.  As such, it is a conspicuous empirical signal of the almost pure NG-boson character of the pion, marking the near perfect expression of Eq.\,\eqref{EasyOne} in this almost-massless state as compared to the incomplete cancellation in the $s$-quark-containing kaon.

There are two issues here.
First, again, there is only one forty-year-old measurement of ${\mathpzc u}^K(x)/{\mathpzc u}^\pi(x)$ \cite{Badier:1980jq}; hence, agreement between this experimental data and the prediction in Ref.\,\cite{Chen:2016sno} \emph{might} be accidental.  A modern measurement is essential.

\begin{figure}[t]
\vspace*{-3.2ex}

\includegraphics[clip, width=0.44\textwidth]{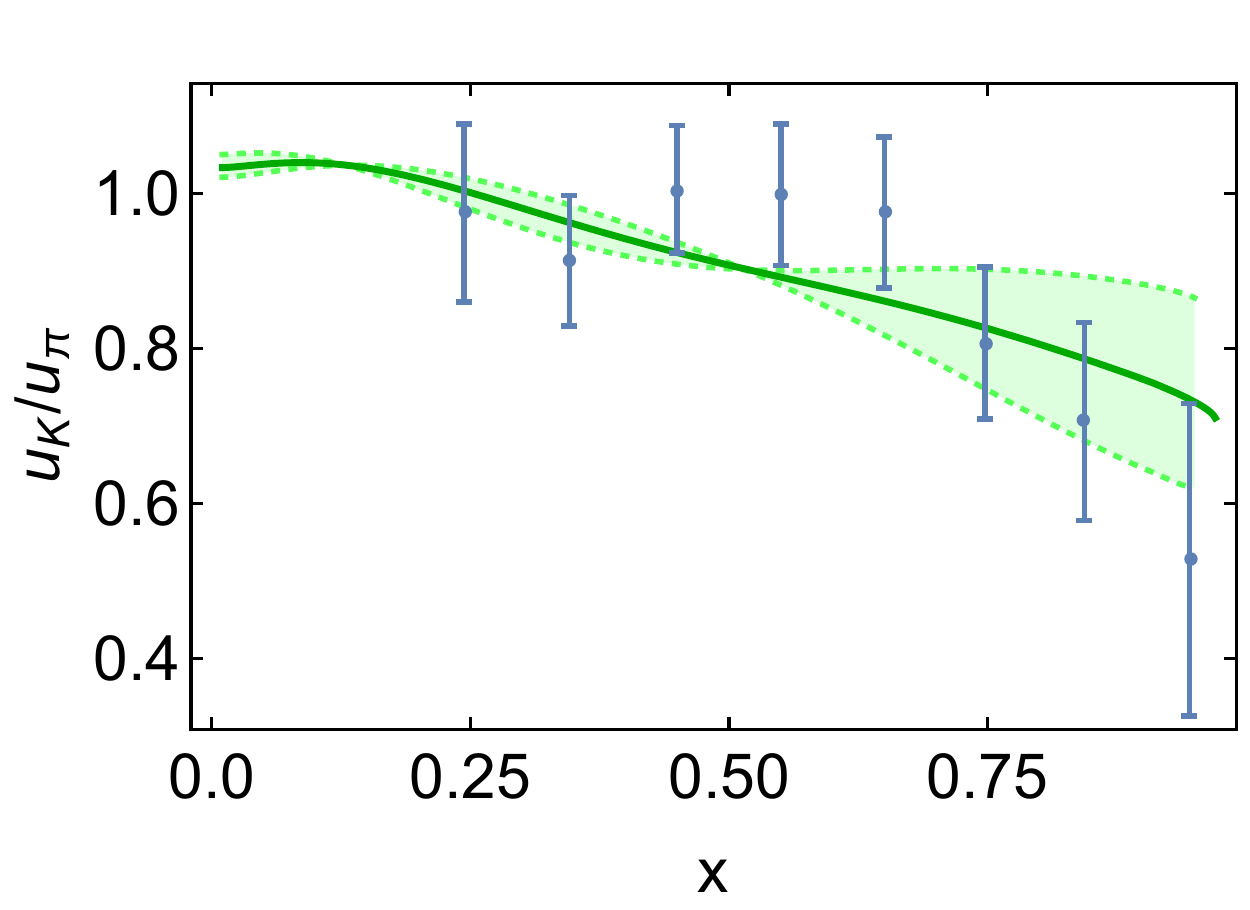}
\caption{\label{FigRatio}
${\mathpzc u}^K(x;\zeta_5)/{\mathpzc u}^\pi(x;\zeta_5)$ computed from the PDFs in Fig.\,\ref{FigPDFs}.  Uncertainty in the kaon PDA, Fig.\,\ref{FigPDAs}, is propagated through the PDF into the behaviour of this ratio on $x\gtrsim 0.5$.  These differences will also be expressed in the sea and glue distributions.
}
\end{figure}

Second, Ref.\,\cite{Chen:2016sno} employed algebraic models for the elements needed to compute the structure function, \emph{i.e}.\ dressed-quark propagators, meson Bethe-Salpeter amplitudes and dressed-photon-quark vertex.  The analysis should thus be repeated using the most sophisticated description of quark+antiquark scattering that is now available, \emph{viz}.\ the DCSB-improved (DB) quark+antiquark scattering kernel \cite{Chang:2009zb, Binosi:2014aea}.  It should also incorporate the connection between $\zeta_H$ and the RGI PI charge, Fig.\,\ref{Figwidehatalpha}, developed in Refs.\,\cite{Ding:2019qlr, Ding:2019lwe, Cui:2019dwv} and the expression of this relationship in DGLAP evolution.

One can embark upon this theory improvement by using Eq.\,\eqref{PDFeqPDA2} and DB-results for the $\pi$- and $K$-meson PDAs.  Fig.\,\ref{FigPDAs} depicts modern reappraisals of these functions, which yield the PDFs in Fig.\,\ref{FigPDFs}.  Working with the PDFs depicted, one obtains the results for ${\mathpzc u}^K(x;\zeta_5)/{\mathpzc u}^\pi(x;\zeta_5)$ drawn in Fig.\,\ref{FigRatio}.  Evidently, the uncertainty existing in the kaon PDA, Fig.\,\ref{FigPDAs}, is expressed in the behaviour of this ratio on $x\gtrsim 0.5$: a broader kaon PDA yields a ratio closer to unity at $x=1$.  Such differences must also manifest themselves in results for the kaon's sea and glue distributions; and the calculation of these functions, following the procedure in Refs.\,\cite{Ding:2019qlr, Ding:2019lwe}, must now be given high priority.  Here the use of mass-dependent DGLAP splitting functions will be crucial.
A crude estimate, based on the ratio of dressed-quark mass functions at infrared momenta, which controls many EHM-induced violations of SU$(3)$-flavour symmetry, suggests such mass-corrections will lead to a form of $\bar {\mathpzc s}^K(x;\zeta_5)$ like that sketched in Fig.\,\ref{FigPDFs}, which yields
\begin{equation}
\label{MassDependent}
\langle x  [{\mathpzc u}^K (x;\zeta_5)+\bar{\mathpzc s}^K(x;\zeta_5)]\rangle /[2 \langle {\mathpzc u}^\pi (x;\zeta_5)\rangle]\sim 1.2\,,
\end{equation}
with $\langle x_{\rm \zeta_5}\rangle_{\rm glue}^K \sim 0.84 \langle x_{\rm \zeta_5}\rangle_{\rm glue}^\pi$,
$\langle x_{\rm \zeta_5}\rangle_{\rm sea}^K \sim 0.96 \langle x_{\rm \zeta_5}\rangle_{\rm sea}^\pi$.  However,  detailed calculations are necessary to improve these simple estimates.

\section{Concluding Remarks}
Discovery of the Higgs boson was a watershed in modern physics.  Now that science has located the source for $\lesssim 2$\% of the mass of visible matter, the focus can shift to searching for the origin of the remaining $\gtrsim 98$\%.  The mechanism at work here must be capable of simultaneously generating the 1\,GeV mass-scale associated with the nucleon and ensuring that this mass-scale is completely hidden in the chiral-limit pion, which must be massless.

This contribution adopts the view that a single renormalisation group invariant (RGI) mass-scale, $m_0$, emerges from strong interactions within the Standard Model and all mass-dimensioned quantities derive their existence and values from $m_0$.  Furthermore, that $m_0$ is the value of the running gluon mass in the infrared.  A combination of the best available continuum and lattice analyses yields $m_0= 0.43(1)\,$GeV.  This value is roughly half the nucleon mass, $m_N$, even though the value of $m_N$ played no direct role in determining it.

The existence of $m_0 \approx m_N/2$ enables an infrared completion of QCD because it provides for a unique mathematical definition of a RGI process-independent (PI) running coupling, $\hat\alpha(k^2)$, which is a smooth function of spacelike momentum-transfers and saturates to a finite value at $k^2=0$: $\hat\alpha(0)/\pi=0.97(4)$  [Fig.\,\ref{Figwidehatalpha}].   There is no Landau pole.  With this appearance of an infrared stable fixed point and the associated behaviour of $\hat\alpha(k^2)$, QCD is seen to be approximately scale-invariant on $0<k^2\lesssim m_0^2$: running practically stops within this conformal window.

The emergence of $m_0$ expresses itself in all sectors of the strong interaction.  For instance, $\hat\alpha(k^2)$ controls the kernel of QCD's gap equation; and with the predicted behaviour of the coupling, the gap equation returns a momentum-dependent solution for the dressed-quark mass, $M(k^2)$, with $M(0) \approx m_N/3$ in the absence of a Higgs mechanism.  Hence, QCD dynamics generates \emph{mass from nothing} in both the gauge and matter sectors.\footnote{Notably, the distinction between gauge and matter sectors is lost if hybrid and/or glueball states are found; and there is every reason to expect they will be.}  These outcomes are the basis for the emergence of hadronic mass (EHM).

It is worth highlighting here that confinement is one word that means many things, \emph{inter alia}: the only measurable asymptotic states are colour-singlets; observable states are colour neutral over a finite, nonzero spacetime volume; and the length scale characterising this volume plays a key r\^ole in defining what is meant by an asymptotic state.  Evidently, therefore, empirical confinement is impossible without the existence of a mass-scale.  Hence, confinement and EHM are tightly linked, perhaps inextricably.  This connection is highlighted by the fact that hadrons are characterised by electromagnetic radii $r \sim 1/m_0 \approx 0.5\,$fm.   It should further be recorded that owing to the emergence of running masses for the gluon and quark, the analytic structure of the associated Schwinger functions changes dramatically; and it can therefrom be argued that parton fragmentation functions provide a clean, empirically-accessible expression of confinement [Sec.\,\ref{EmpConfinement}].

A striking feature of QCD is that EHM is hidden in the pion mass: $m_\pi\equiv 0$ in the absence of a Higgs mechanism, no matter how large is the value of $m_N$.  Although it is hidden here, EHM is not dormant.  Indeed, the pion is massless because the EHM mechanism enforces complete cancellations between $2\times$\,one-body and $1\times\,$two-body dressing effects [Sec.\,\ref{NGPion}].  In other channels, the cancellations are neither complete nor tuned; hence, all other hadron masses take values commensurate with $m_0$.  Moreover, for every hadron, EHM introduces a dynamical enhancement factor that multiplies the contribution from the Higgs-generated current-mass term to that hadron's mass.  For the nucleon, this factor is almost one order-of-magnitude.


In our view, the emergence of $m_0$ and its expression in the properties of $\hat\alpha(k^2)$ are the keys to explaining EHM.  A primary goal, therefore, is to elucidate those of their manifold consequences which are best suited to empirical observation.  This potential is expressed in all areas of hadro-particle physics.  Herein, we concentrated on signature features of pion and kaon parton distributions (PDFs) [Sec.\,\ref{ObservableEHM}].

In connection with such PDFs, we highlighted a thirty-year controversy, which began with Ref.\,\cite{Conway:1989fs}.  Namely, QCD makes a clean prediction for the large-$x$ behaviour of the PDF associated with spin-zero hadrons at the hadronic scale, ${\mathpzc q}(x;\zeta_H=m_0)$ [Eq.\,\eqref{PDFQCD}].
%
The hadronic scale, $\zeta_H$, is not accessible in Drell-Yan (DY) and deep inelastic scattering (DIS) processes because certain kinematic conditions need to be met in order for the data to be interpreted in terms of distribution functions.  In these circumstances, the QCD prediction translates into the following statement:
\begin{equation}
\label{dpiDY}
{\mathpzc q}(x\simeq 1;\zeta>\zeta_H) \propto (1-x)^{\beta}\,,\; \beta >2\,.
\end{equation}
Modern QCD-connected theory conforms with Eq.\,\eqref{dpiDY}; but many models do not.  

Amongst extant analyses of data relating to the pion's valence-quark PDF, only Ref.\,\cite{Aicher:2010cb} employs a fully consistent next-to-leading-order analysis, including threshold resummation; and only this study yields a result for ${\mathpzc q}^\pi(x;\zeta)$ which agrees with Eq.\,\eqref{dpiDY}.
All other analyses ignore threshold resummation, producing $\beta <2$, in conflict with Eq.\,\eqref{dpiDY} but consistent with some of the popular models.
A mismatch with QCD for the phenomenologically determined valence quark distribution means that contingent results for the pion's sea and glue distributions are potentially unsound [Fig.\,\ref{FigcfJAM}].  This is an issue because reliable empirically derived information about the pion's sea and glue content is highly desirable.

These remarks highlight that NG modes are far more complex than is typically thought.  They are not pointlike; they are intimately connected with the origin of mass; and they probably play an essential part in any answer to the question of gluon and quark confinement in the physical Universe [Sec.\,\ref{EmpConfinement}].  Indeed, the internal structure of NG modes is intricate; and that structure provides the clearest window onto EHM, \emph{e.g}.\ the EHM mechanism may entail that the gluon content within Nature's only near-pure Nambu-Goldstone mode, the pion, is significantly larger than that in any other hadron.  This notion can be tested in comparisons between measurements of pion and kaon parton distribution amplitudes and functions.

Pions and kaons are critical to the formation of everything from nucleons, to nuclei, and on to neutron stars.  Hence, new-era experiments capable of validating the predictions described herein are of the highest priority.  With validation, an entire chapter of the Standard Model, whose writing began more than eighty years ago \cite{Yukawa:1935xg}, can be completed and closed with elucidation of the structural details of the Standard Model's only NG modes.

The hunt for an understanding of more-than 98\% of the visible mass in the Universe has been underway for more than forty years \cite{Lane:1974he, Politzer:1976tv, Pagels:1978ba, Cornwall:1981zr, Higashijima:1983gx,  Munczek:1983dx, Fomin:1984tv, Cahill:1985mh}.  Now, however, with the Higgs boson catalogued, the curtain has been swept away and a wider community can see the need for this search.

The key to understanding the origin and properties of the vast bulk of all known matter is the strong interaction sector of the Standard Model.  The current paradigm is QCD, plausibly the only mathematically well defined four-dimensional theory that science has ever produced.  Hence, the goal is to reveal the content of strong-QCD.  In working toward this, no one approach is sufficient.  Progress and insights are being delivered by an amalgam of experiment, phenomenology and theory; and continued exploitation of the synergies between these efforts is essential if science is to capitalise on new opportunities provided by existing and planned facilities.

\smallskip

%
\noindent{\sf \textbf{Acknowledgments}}.
This summary of our contributions to the ``40th Max Born Symposium'', held in the Institute of Theoretical Physics at the University of Wroc{\l}aw, Poland, 9-12 October 2019, is based on results obtained through collaborations with many people, to all of whom we are greatly indebted.
We are also grateful to the organisers of this Symposium, for their assistance, kindness and hospitality;
and to
V.~Andrieux, W.-C.~Chang, O.~Denisov, J.~Friedrich, W.~Melnitchouk, W.-D.~Novak, S.~Platchkov, M.~Quaresma and C.~Quintans --
from whom we received valuable input during preparation of this manuscript.
Work and participation supported in part by:
Jiangsu Province \emph{Hundred Talents Plan for Professionals};
and Polish Academy of Sciences.
%



\end{document}